# High Accuracy and Cost-Saving Active Learning: 3D WD-UNet for Airway Segmentation

Shiyi Wang
National Heart and Lung Institute, Imperial College London
London, United Kingdom
s.wang22@imperial.ac.uk

Yang Nan
National Heart and Lung Institute, Imperial College London
London, United Kingdom
y.nan20@imperial.ac.uk

Walsh Simon L F
Royal Brompton Hospital
National Heart and Lung Institute, Imperial College London
London, United Kingdom
s.walsh@imperial.ac.uk

Guang Yang
Cardiovascular Research Centre, Royal Brompton Hospital
National Heart and Lung Institute, Imperial College London
London, United Kingdom
g.yang@imperial.ac.uk

*Abstract*—We propose a novel Deep Active Learning (DeepAL) model – 3D Wasserstein Discriminative UNet (WD-UNet) for reducing the annotation effort of medical 3D Computed Tomography (CT) segmentation. The proposed WD-UNet learns in a semi-supervised way and accelerates learning convergence to meet or exceed the prediction metrics of supervised learning models. Our method can be embedded with different Active Learning (AL) strategies and different network structures. The model is evaluated on 3D lung airway CT scans for medical segmentation and show that the use of uncertainty metric, which is parametrized as an input of query strategy, leads to more accurate prediction results than some state-of-the-art Deep Learning (DL) supervised models, e.g., 3D UNet and 3D CEUNet. Compared to the above supervised DL methods, our WD-UNet not only saves the cost of annotation for radiologists but also saves computational resources. WD-UNet uses a limited amount of annotated data (35% of the total) to achieve better predictive metrics with a more efficient deep learning model algorithm.

## I. Introduction

### A. Active Learning

Deep learning (DL) has demonstrated exceptional performance in numerous medical image segmentation tasks, including lung disease diagnosis and organ/cortex segmentation [1], [2]. Despite the potential of DL models to revolutionize clinical management, the acquisition of expert-annotated data to train these models remains a costly, time-consuming, and subjective process, particularly for large-scale 3D medical segmentation datasets with dense collective elements. Active learning (AL), on the other hand, is a strategy that selects and labels the most informative samples during an interactive querying cycle (see Figure 1). AL enables more efficient use of resources by selectively choosing the most valuable data samples for annotation, thereby reducing the need for laborious and costly annotation efforts [3].

### B. Deep Active Learning

Deep Active Learning (DeepAL) combines active learning with deep learning models to identify informative data samples for annotation. In Figure 2, (1) Initially, deep learning models are pre-trained on a small labeled dataset. (2) Then, the models are iteratively trained by selecting the most informative or uncertain samples using query strategies such as uncertainty or least confidence. (3)(4) These selected samples are labeled by experts and added to the training set, and the model is retrained. This process continues until the desired model performance is achieved or the annotation budget is exhausted. DeepAL benefits from the powerful information extraction and high-dimensional image processing capabilities of deep learning models, while leveraging the efficiency of active learning to reduce annotation costs [4]. Compared to traditional low-shot segmentation schemes and co-segmentation methods, our proposed WD-UNet model has the advantages of independence from pre-training quality and strong generalization and stability.

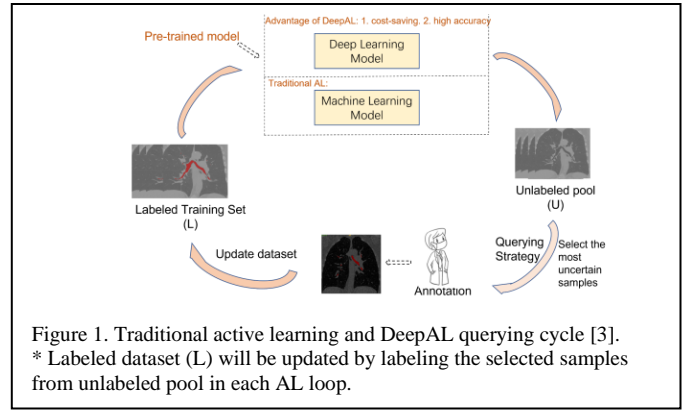

Figure 1. Traditional active learning and DeepAL querying cycle [3].
\* Labeled dataset (L) will be updated by labeling the selected samples from unlabeled pool in each AL loop.

## II. Methods

### A. Dataset.

A total of 108 cases from a mix of the EXACT09 dataset [5] and LIDC-IDRI dataset [6] of 3D HRCT data were used for training and testing (divided as 72, 18,18). Each case contains a pair of the original images and ground truth. Since the full 3D CT image is too large to be used as input to the deep convolutional network (DNN), the data needs to be pre-processed and sliced to a 3D patch of size [128,96,144]. After data pre-processing, there are 1015 patches in total, 406 patches are used for initial training (approximately 36 cases), 406 patches are unlabeled images (approximately 36 cases), and 203 patches (approximately 18 cases) are used for testing accuracy after each querying cycle of DeepAL. The rest 18 cases are used for inference. All the data employed for inference purposes is distinct from the datasets used for training and testing.

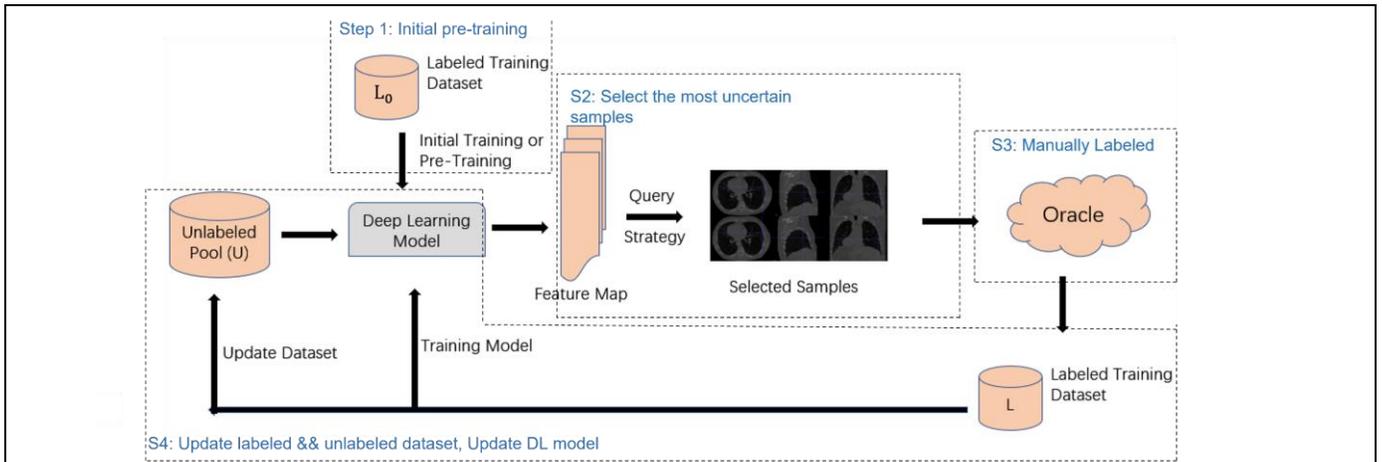

Figure 2. Our framework describes a typical deep active learning structure. Some parameters will be defined by deep learning model based on initial training or pre-training on the small scale of labeled dataset L0. Then the features are extracted by the previous deep learning model from the unlabeled pool (U). The related query strategy can select representative samples and form them in the labeled training dataset (L). The new labeled dataset can be used to update U and fed to deep learning model for the new iteration of training. This processing loop is continuing, until the expected the label budget and pre-defined termination conditions are achieved. *L0 is the initial training dataset, L is the updated dataset from each AL loop. After each active learning loop, the data in the unlabeled set U will be labeled and transferred to the labeled set L.

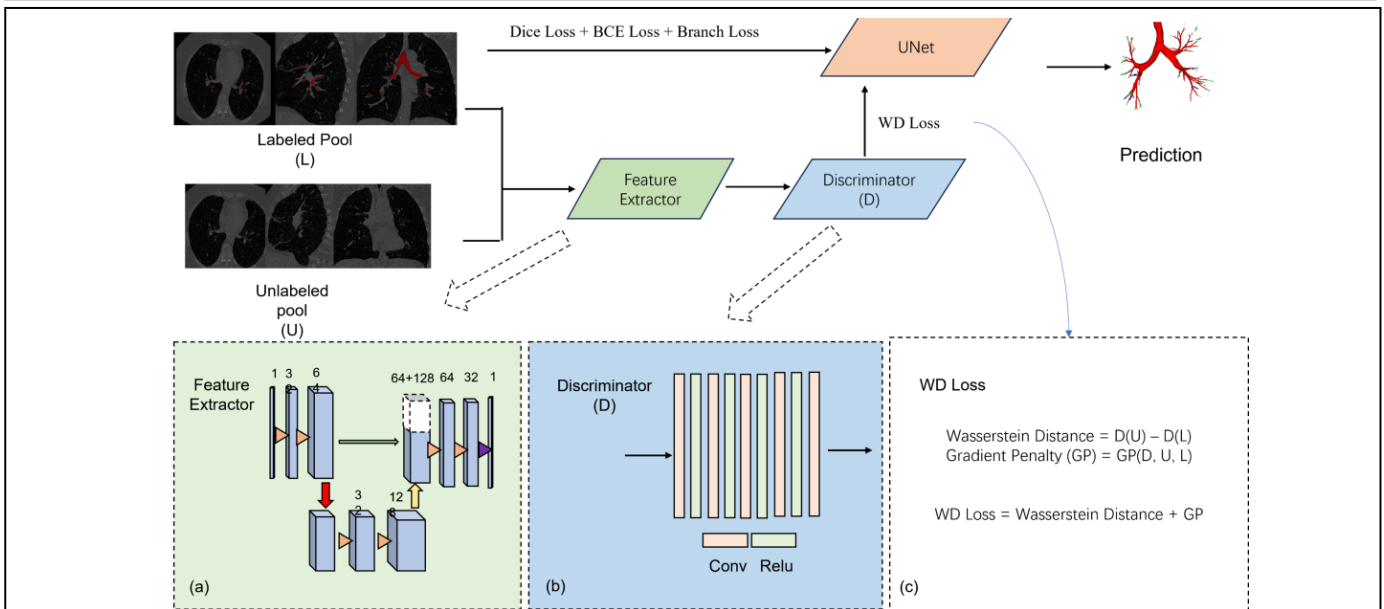

Figure 3. The whole training procedure of WD-UNet. *The explanation of Branch Loss, Discriminator, Feature Extractor, Wasserstein Distance and Gradient Penalty can be found in Methods section. WD Loss (Wasserstein Discriminator Loss) is composed of Wasserstein Distance and Gradient Penalty (GP), and the total loss should contain Dice, BCE, Branch and WD Loss (a) shows the details of Feature Extractor which can be seen as a simple UNet structure with one down sample, an up sample and a skip connection. It aims to rapidly extract feature representations from labeled and unlabeled images. (b) shows the structure of a Discriminator with 5 Conv layers and 4 Relu. (c) computes the diversity (differences) between unlabeled data and labelled data, enabling higher transportation costs under Wasserstein distance and improved gradient information to ensure model stability and diversity.

*B. Our Benchmark*

To establish a comparative analysis with the DeepAL approaches (semi-supervised), we introduce two supervised baseline models, namely the 3D UNet [7][8] and the CEUNet. The intention behind this is to enable a performance comparison between our proposed WD-UNet algorithm and the supervised models. In our forthcoming work, we propose a variant of the U-Net architecture called CEUNet, which incorporates a dense atrous convolution (DAC) module and a Residual Multinuclear Pooling (RMP) module between the encoder and decoder components [9]. This modification allows CEUNet to utilize convolutions with varying dilation rates, enabling the extraction of features from objects of different sizes and producing feature maps of varying resolutions.

*C. 3D Wasserstein Discriminative UNet*

The proposed WD-UNet model incorporates a deep learning (DL) model known as 3DUNet. This DL model can be retrained in each active learning (AL) querying cycle utilizing query strategies to identify the most informative data for annotation. Consequently, this iterative process updates the model until it achieves the desired outcome. For the training procedure, in Figure 3. WD-UNet has two main steps, (1) The labeled data is

**Algorithm 1** Gradient Penalty Approach

---
**Require:** Critic network, unlabeled data, labeled data

**Ensure:** Penalty value

1: alpha = torch.rand()
2: interpolates = (alpha, unlabeled data, labeled data)
3: preds = critic(interpolates)
4: gradients = torch.autograd.grad(preds, interpolates)
5: **if** gradient.norm > max_norm:
   clip_coef = max_norm / gradient.norm
   gradients = gradients * clip_coef.
6: **end for**
7: penalty = ((gradient.norm - 1) ** 2).mean()
8: **returns** penalty

---

fed to the model for training and calculating the Dice, BCE, and Branch Loss. (2) The labeled data and unlabeled data are sequentially fed into the feature extractor and the Discriminator. This process generates label_out and unlabel_out, which are then utilized to compute the Wasserstein distance between the two. Subsequently, label_out, unlabel_out, and the Discriminator are collectively passed into the Gradient Penalty function to calculate the penalty, which is obtained by evaluating the mean of the squared difference between the gradient norm and 1. Finally, the penalty is added to the Wasserstein distance, yielding the WD Loss. Finally, these 5 loss functions are integrated to optimize the WD-UNet model.

**Wasserstein distance**, also known as Earth Mover's Distance (EMD) [10], [11], provides an effective measure of dissimilarity between two distributions, particularly for tasks involving image generation or image transformation. Compared to other distance metrics such as Kullback-Leibler (KL) divergence or Jensen-Shannon (JS) divergence, Wasserstein distance can more accurately quantify the differences between two distributions, even in complex scenarios with mode collapse, while providing better gradient information. The principle behind Wasserstein distance is based on the theory of Optimal Transport in optimization problems. Its core idea involves considering the minimum cost required to transform one distribution into another. In our proposed WD-UNet, however, Wasserstein distance is employed to differentiate between unlabeled and labeled empirical distributions, with their discrepancy serving as one of the loss functions for WD-UNet. This allows the model to select diverse training data. The unlabeled and labeled empirical distributions are computed by a defined discriminator. Regarding uncertainty, we aim to identify samples with lower predictive confidence (i.e., higher uncertainty). As for diversity, we seek to find unlabeled batches with higher transportation costs under Wasserstein distance compared to the labeled set (i.e., dissimilar to the current labeled samples), which has been shown to be a good measure of diversity.

**Gradient penalty** is a function used to compute the gradient penalty in training the discriminator network for computing the Wasserstein distance. It takes as input the discriminator network, unlabeled data, and labeled data. The gradient penalty is employed to encourage the discriminator to maintain gradients close to 1 for interpolated samples, thus improving the training effectiveness and stability of the model. Additionally, this module calculates the mean of the squared difference between the gradient norm and 1, referred to as the penalty, which serves as both a gradient penalty value and one of the loss functions for updating the model.

**Feature extractor** is a module used to extract meaningful features from raw data. Its main purpose is to transform the raw data into a discriminative representation that supports the training of active learning. By performing feature extraction on both unlabeled and labeled data, the following effects can be achieved: (1) Extracting lower-dimensional feature representations that capture key information in the data, thereby facilitating better understanding and differentiation of different classes or attributes when computing the Wasserstein distance in the Discriminator. (2) Removing irrelevant information to improve the efficiency and generalization ability of the model.

**Branch Loss**: Branch_label represents the index of each branch. Each branch can be calculated and labeled by the parent-children relationship. Due to the availability of tree parsing, it is possible to compute the branches for each segment. The intersection is obtained by multiplying the predicted branch for each segment with the corresponding ground truth (GT) branch and summing the results. The denominator is obtained by summing the branches of the ground truth for each segment. Consequently, the branch loss can be expressed as follows:

$$L_{branch\_loss} = 1 - \frac{\sum Pred_{branch\_label} * GT_{branch_{label}} + smooth}{\sum GT_{branch\_label} + smooth} \quad (1)$$

At WD-UNet, a loss function has been devised that capitalizes on the specificity of lung tracheal segmentation to optimize the segmentation outcomes of a deep learning model. This loss function enhances the segmentation accuracy by refining the detail-to-structure aspects of the model's performance. Unlike conventional medical segmentation deep learning algorithms, which rely on the Dice Loss (dice similarity coefficient) and BCE (binary cross entropy) Loss weights, the proposed training loss function leverages the unique characteristics of the trachea to compute the Branch Loss of tracheal branches. The branch loss measures how well the predicted airway centerline matches the ground truth airway centerline, with a value of 0 indicating a perfect match and a value of 1 indicating no overlap between the two. The smooth factor is added to avoid numerical instability caused by division by zero. The total loss function can be illustrated as Equation 2:

$$L_{WD} = L_{Wasserstein\ Distance} + L_{Penalty}$$
$$L_{total} = L_{Dice}^{w} + L_{BCE}^{w} + L_{Branch} + L_{WD} \quad (2)$$

**Query Strategy.** First, the indices and data of the unlabeled samples are obtained. Next, the predict_prob method is used to predict the probabilities of the unlabeled data, resulting in probability values. Two scores need to be calculated: (1) the uncertainty score, which is obtained by weighting the upper bounds of the L2 norm and the L1 norm. (2) The discriminative score of the predicted output of the unlabeled data based on the Discriminator. Finally, the total score is computed by subtracting the product of the discriminative score from the product of the uncertainty score and the selection parameter.

TABLE I. Evaluation metrics of the airway segmentation results between supervised models and semi-supervised models with different ratios of training data. * Dice Similarity Coefficient (DSC), branch score (BD), tree detected ratio (TD), Intersection over Union (IoU). In the table, the bold text indicates the model performance that requires specific emphasis and discussion in the paper.

| Evaluation Metrics | Our Models | | | | | |
|---|---|---|---|---|---|---|
| | Supervised model | | Semi-supervised models | | | |
| | U-Net | CEUNet | UUNet (15% training data) | UUNet (35% training data) | UUNet (55% training data) | UUNet (75% training data) |
| DSC | 0.8954±0.0282 | 0.8990±0.0293 | 0.8930±0.0370 | **0.8381±0.0972** | 0.8958±0.0366 | 0.8958±0.0345 |
| Precision | 0.8905±0.0448 | 0.9115±0.0418 | 0.9013±0.0462 | **0.8957±0.0465** | 0.9013±0.0507 | 0.9080±0.0420 |
| TD | 0.8118±0.0467 | 0.8178±0.0482 | 0.8216±0.1208 | **0.8691±0.1082** | 0.8784±0.0677 | **0.8969±0.0754** |
| BD | 0.8698±0.0777 | 0.8421±0.0983 | 0.7524±0.1704 | **0.8243±0.125** | 0.8456±0.1046 | 0.8630±0.1000 |
| IoU | 0.8078±0.0548 | 0.8150±0.0560 | 0.8002±0.0641 | **0.8092±0.0540** | 0.8132±0.0571 | 0.8078±0.0291 |

TABLE II. Evaluation metrics of the airway segmentation results between 2 supervised models and 4 semi-supervised models. LC-UNet is based on least confidence [12], UUNet is based on uncertainty estimation which is obtained entropy sampling[3], RS-UNet is based on random sampling [13].

| Evaluation Metrics | Our Models | | | | | |
|---|---|---|---|---|---|---|
| | Supervised model | | Semi-supervised models | | | |
| | U-Net | CEUNet | LC-UNet | UUNet | RS-UNet | Proposed WD-UNet |
| DSC | 0.8954±0.0282 | 0.8990±0.0293 | 0.8965±0.0318 | 0.8381±0.0972 | 0.8907±0.0318 | **0.9008±0.0360** |
| Precision | 0.8905±0.0448 | 0.9115±0.0418 | 0.9084±0.0460 | 0.8957±0.0465 | 0.8882±0.0496 | **0.9088±0.0459** |
| TD | 0.8118±0.0467 | 0.8178±0.0482 | 0.8597±0.0990 | 0.8691±0.1082 | 0.8881±0.0706 | **0.8928±0.0834** |
| BD | 0.8698±0.0777 | 0.8421±0.0983 | 0.8165±0.1200 | 0.8243±0.125 | 0.8476±0.0956 | **0.8485±0.1214** |
| IoU | 0.8078±0.0548 | 0.8150±0.0560 | 0.8088±0.0499 | 0.8092±0.0540 | 0.8044±0.0527 | **0.8139±0.0582** |

### III. RESULTS AND DISCUSSIONS

#### A. Qualitative Findings

Combining the findings from Table I and Table II, it is observed that the UUNet (based on entropy sampling) with the least computational complexity is selected from Table I. The UUNet model is trained using different proportions of training data (15%, 35%, 55%, and 75%). It is found that the UUNet achieves impressive performance when trained with 35% of the training data, exhibiting comparable performance to the two supervised models (UNet and CEUNet). In fact, it even surpasses UNet in the tree detected ratio (TD), with values of 0.8691±0.1082 and 0.8118±0.0467, respectively. In Table I, it can be observed that as the proportion of training data increases from 15% to 75%, the model initially achieves high values of DSC 0.8930 at 15%. However, the DSC drops to 0.8381 at 35%. Interestingly, the DSC then improves again at 55% and 75%. Based on these findings, it is reasonable to infer that DeepAL methods perform well on smaller datasets, show a slight decline as diversity increases (due to out-of-scope samples), and eventually stabilize with improved DSC owing to the strong generalization ability of the DeepAL model. In Table II, the LC-UNet, UUNet, RS-UNet, and the proposed WD-UNet model, which were implemented by us, were trained using 35% of the training data. A comparison was made between these models and the two supervised models. Experimental results revealed that the WD-UNet model outperformed the other three DeepAL models in terms of all five performance metrics. Furthermore, it also surpassed the UNet and CEUNet models in terms of the DSC and tree detected ratio (TD).

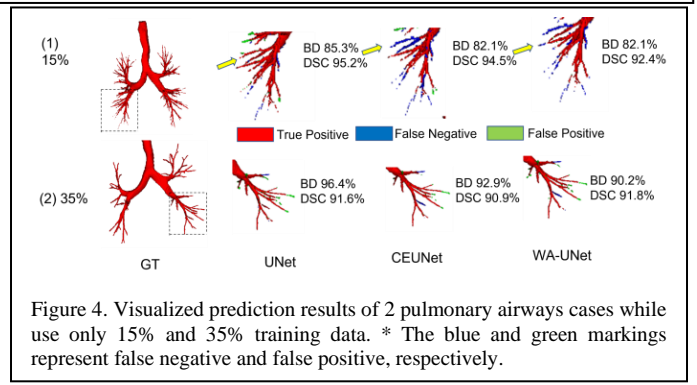

Figure 4. Visualized prediction results of 2 pulmonary airways cases while use only 15% and 35% training data. * The blue and green markings represent false negative and false positive, respectively.

#### B. Quantitative Findings

When using 15% of the training data, Figure 4 visually demonstrates that WD-UNet exhibits weaker prediction capability for the distal trachea compared to UNet (with more false negative regions), but is closer to CEUNet. However, when utilizing 35% of the training data, the prediction capability of the WD-UNet model significantly improves and approaches that of UNet (with fewer false negative regions). The green regions represent false positives, indicating that while some prediction errors may occur, they primarily indicate the model's ability to detect subtle branches in the distal trachea that may not be visible to radiology experts.

## IV. Conclusion

In summary, DeepAL achieves impressive results even with only 35% of the training data, and WD-UNet's false negative is steadily decreasing as the proportion of training data increases and its ability to predict fine branches in the distal part of the trachea improves. When applied to clinical practice, WD-UNet demonstrates a competitive advantage because it can flexibly balance performance and cost by determining the trade-off between evaluation metrics and the amount of training data required. Additionally, WD-UNet demonstrates superior performance compared to other state-of-the-art deep methods reproduced in this study (TABLE II), owing to its diversity prediction and stability.